\documentclass[11pt]{article}
\pdfoutput=1

\usepackage[T1]{fontenc}              % T1 font encoding as default
\usepackage[latin9]{inputenc}         % ISO-8859-15 (Latin 9) text

\usepackage[a4paper]{geometry}        % DIN A4 paper

\usepackage{url}

\usepackage{booktabs}

\usepackage{graphicx} % WE USE PDFLATEX
\usepackage{amssymb,amsfonts,amsmath}

\usepackage{natbib}
\usepackage{mathpazo}
\usepackage{microtype}

\usepackage{color}
\newcommand\bdelta{\boldsymbol \delta}
\newcommand\bI{\boldsymbol I}
\newcommand\bM{\boldsymbol M}
\newcommand\bmu{\boldsymbol \mu}
\newcommand\bomega{\boldsymbol \omega}

\newcommand\bP{\boldsymbol P}
\newcommand\bR{\boldsymbol R}
\newcommand\bS{\boldsymbol S}
\newcommand\bSigma{\boldsymbol \Sigma}
\newcommand\btau{\boldsymbol \tau}

\newcommand\bV{\boldsymbol V}
\newcommand\bX{\boldsymbol X}
\newcommand\bx{\boldsymbol x}

\newcommand\bb{\boldsymbol b}

\newcommand\expect{\text{E}}

\newcommand\var{\text{Var}}
\newcommand\cov{\text{Cov}}
\newcommand\corr{\text{Corr}}

\newcommand{\figcite}[1]{Fig.~\textbf{\ref{#1}}}
\newcommand{\eqcite}[1]{Eq.~\textbf{\ref{#1}}}
\newcommand{\tabcite}[1]{Tab.~\textbf{\ref{#1}}}

\begin{document}

\date{18 July 2011}

\title{
High-Dimensional Regression and Variable Selection Using CAR Scores
}

\author{Verena Zuber 
      \thanks{Institute for Medical Informatics,
      Statistics and Epidemiology,
      University of Leipzig,
      H\"artelstr. 16--18,
      D-04107 Leipzig, Germany} 
{} and
 Korbinian Strimmer \footnotemark[1]
}

\maketitle

\begin{center}
Statistical Applications in Genetics and Molecular Biology {\bf 10}: 34 (2011).
\end{center}
\vspace{2cm}

\begin{abstract}

    Variable selection is a difficult problem that is particularly
challenging in the analysis of high-dimensional genomic data. Here, we 
introduce the CAR score, a novel and highly effective criterion for 
variable ranking in linear regression based on Mahalanobis-decorrelation 
of the explanatory variables. The CAR score provides a canonical ordering 
that encourages grouping of correlated predictors and down-weights 
antagonistic variables.  It decomposes the proportion of variance 
explained and it is an intermediate between marginal correlation and
the standardized regression coefficient. As a population quantity, 
any preferred inference scheme can be applied for its estimation.
Using simulations we demonstrate that variable selection by CAR scores 
is very effective and yields prediction errors and true and false 
positive rates that compare favorably with modern regression techniques 
such as elastic net and boosting. We  illustrate our approach by 
analyzing data concerned with diabetes progression and with the effect 
of aging on gene expression in the human brain. 
The R package "care" implementing CAR score regression is available from CRAN.

\end{abstract}

\newpage

\section{Introduction}

Variable selection in the linear model is a classic statistical problem 
\citep{Geo2000}. The last decade with its immense technological advances
especially in the life sciences has
revitalized interest in model selection in the context
of the analysis of high-dimensional data sets \citep{FL2010}.
In particular, the advent of large-scale genomic data sets
has greatly stimulated the development of 
novel techniques for regularized inference from small samples \citep[e.g.][]{HTF09}. 

Correspondingly, many regularized regression  approaches that automatically
perform model selection  
have been introduced with great success, such as
least angle regression \citep{EHJT2004}, elastic net \citep{ZH2005},
the structured elastic net \citep{LL08}, OSCAR \citep{BR08}, 
the Bayesian elastic net \citep{LL2010}, and the random lasso \citep{WN+2011}.
By construction, in all these methods variable selection is tightly 
linked with a specific inference procedure, typically of Bayesian flavor or
using a variant of penalized maximum likelihood.

Here, we offer an alternative view on model selection in the linear
model that operates on the population level and 
is not tied to a particular estimation paradigm.
We suggest that variable ranking, aggregation and
selection in the linear model is best understood and conducted  
on the level of standardized, Mahalanobis-decorrelated predictors. 
Specifically, we propose  CAR scores, defined as the marginal correlations adjusted for 
correlation among explanatory variables, as a natural variable
importance criterion.
This quantity emerges from a predictive view of the linear model
and leads to a simple  additive 
decomposition of the proportion of explained variance
and to a canonical ordering of the explanatory variables.
By comparison of CAR scores with various other
variable selection and regression approaches, including elastic net, lasso
and boosting, we show that CAR scores, despite their simplicity, are capable
of effective model selection both in  small and in large sample situations.

The remainder of the paper is organized as follows.  First, we revisit the
linear model from a predictive population-based view and briefly review standard
variable selection criteria. Next, we introduce the CAR score and discuss its
theoretical properties. Finally, we conduct extensive computer simulations
as well as data analysis to investigate the practical performance of CAR scores
in high-dimensional regression.

%\newpage
\section{Linear model revisited}

In the following, we recollect basic properties of the linear
regression model from the perspective of the best linear predictor
\citep[e.g.][Chapter 5]{Whi90}.  

\subsection{Setup and notation}

We are interested in modeling the linear relationship between
a metric univariate response variable $Y$ and a vector of predictors
 $\bX = (X_1, \ldots, X_d)^T$.  
We treat both $Y$ and $\bX$ 
as random variables, with means  $\expect(Y) = \mu_Y$ and 
$\expect(\bX)=\bmu$ and (co)-variances $\var(Y) =  \sigma^{2}_Y$, $\var(\bX) = \bSigma$, and 
$\cov(Y, \bX) =\bSigma_{Y \bX} = \expect\left( (Y-\mu_Y) (\bX-\bmu )^T\right) = \bSigma_{\bX Y}^T$.
The matrix $\bSigma$ has dimension 
$d \times d$ and $\bSigma_{Y \bX}$ is of size  $1 \times d$.
With $\bP$ (= capital ``rho'') and $\bP_{Y \bX}$ we denote the correlations among predictors
and the marginal correlations between response and predictors, respectively.
With $\bV = \text{diag}\{\var(X_1), \ldots, \var(X_d)\}$ we decompose
$\bSigma = \bV^{1/2} \bP \bV^{1/2}$ and
$\bSigma_{Y \bX} = \sigma_{Y} \bP_{Y \bX} \bV^{1/2}$.

\subsection{Best linear predictor}

The \emph{best linear predictor} of $Y$ is the linear combination
of the explanatory variables 
\begin{equation}
\label{eq:linpred}
Y^\star =  a + \bb^T \bX   
\end{equation}
that minimizes the mean squared prediction
error $\expect\left((Y-Y^\star)^2\right)$.  This 
is achieved for regression coefficients
\begin{equation}
\label{eq:acoef}
\bb =   \bSigma^{-1}  \, \bSigma_{\bX Y}
\end{equation}
and intercept
\begin{equation}
\label{eq:bcoef}
a = \mu_Y - \bb^T \bmu \,.
\end{equation}
The coefficients  $a$ and $\bb = (b_1, \ldots, b_d)^T$ are \emph{constants}, and not
random variables like $\bX$, $Y$ and $Y^\star$.
The resulting minimal prediction error is
$$
\expect\left((Y-Y^\star)^2\right) = \sigma^2_Y - \bb^T \bSigma \,\bb \,.
$$
Alternatively, the irreducible error may be written  
$\expect\left((Y-Y^\star)^2\right) = \sigma^2_Y \, (1-\Omega^2)$ 
where
$\Omega = \corr(Y, Y^\star)$ and
$$
\Omega^2 = \bP_{Y \bX} \bP^{-1} \bP_{\bX Y}
$$ is the squared 
multiple correlation coefficient. 
Furthermore, $\cov(Y, Y^\star) = \sigma^2_Y  \, \Omega^2$
and $\expect(Y^\star) = \mu_Y$.  The expectation $\expect\left((Y-Y^\star)^2\right)
= \var(Y-Y^\star)$ is also called the \emph{unexplained variance} or
\emph{noise variance}.
Together with the
\emph{explained variance} or  \emph{signal variance}
$\var(Y^\star) = \sigma^2_Y  \, \Omega^2$
it adds up to the \emph{total variance} $\var(Y)=\sigma_Y^2$.
Accordingly,
the \emph{proportion of explained variance} is
$$
\frac{\var(Y^\star)}{ \var(Y) } = \Omega^2 \,,
$$
which indicates that $\Omega^2$ is the central quantity 
for understanding both nominal prediction error and variance
decomposition in the linear model.  The \emph{ratio of
signal variance to noise variance} is 
$$
\frac{\var(Y^\star)}{\var(Y-Y^\star)} = \frac{\Omega^2}{1-\Omega^2} \,.
$$ 
A summary of these relations is given in \tabcite{tab:vardecomp},
along with the empirical error
 decomposition in terms of observed sum of squares.

\begin{table}[t]
\caption{Variance decomposition in terms of squared multiple correlation 
$\Omega^2$ and corresponding empirical sums of squares.}
\begin{center}
\begin{tabular}{llclcl}
\toprule
Level & Total variance & $=$ & unexplained variance & $+$ & explained variance \\
\midrule
\midrule
Population & $\var(Y)$ & $=$  & $\var(Y - Y^\star)$ &  $+$  & $\var(Y^\star)$ \\ 
 & $\sigma^2_Y$ & $=$ & $\sigma^2_Y \, (1-  \Omega^2)$ &  $+$ & $\sigma^2_Y \,  \Omega^2$ \\ 
\midrule
Empirical & $\text{TSS}$  &  $=$ & $\text{RSS}$ &  $+$    & $\text{ESS}$ \\
 &$\sum_{i=1}^n (y_i -\bar{y})^2$ & $=$ & $\sum_{i=1}^n (y_i -\hat y_i)^2$ & $+$ & 
$\sum_{i=1}^n (\hat y_i -\bar{y})^2$ \\
& $\text{d.f.} = n-1$ & & $\text{d.f.} = n-d-1$ & & $\text{d.f.} = d$\\
\bottomrule 
%\multicolumn{6}{c}{Abbreviations: $\bar{y} = \frac{1}{n} \sum_{i=1}^n y_i$; d.f: degrees of freedom; 
%TSS: total sum of squares; RSS: residual sum of squares; ESS: explained sum of squares.}\\
\end{tabular}
Abbreviations: $\bar{y} = \frac{1}{n} \sum_{i=1}^n y_i$; d.f: degrees of freedom; 
TSS: total sum of squares; RSS: residual sum of squares; ESS: explained sum of squares.
\end{center}
\label{tab:vardecomp}
\end{table}

If instead of the optimal parameters $a$ and $\bb$ we employ
$a' = a + \Delta a$ and $\bb' = \bb + \Delta \bb$ 
the minimal mean squared prediction error $\expect\left((Y-Y^\star)^2\right)$
increases by the \emph{model error}
$$
ME(\Delta a, \Delta \bb) = (\Delta \bb)^T \, \bSigma \, \Delta \bb + (\Delta a)^2 \,.
$$
The \emph{relative model error} is the ratio of the model error
and the irreducible error $\expect\left((Y-Y^\star)^2\right)$.

\subsection{Standardized regression equation}

Often, it is convenient to center and standardize the
response and the predictor variables.   With
$Y_{\text{std}} = (Y-\mu_Y)/\sigma_{Y}
$
and 
$
\bX_{\text{std}}  = \bV^{-1/2} (\bX -\bmu )
$
the predictor equation (\eqcite{eq:linpred}) can be written as
\begin{equation}
\label{eq:linpredstd}
Y^\star_{\text{std}} =  ( Y^\star -\mu_Y)/\sigma_{Y} =
  \bb^T_{\text{std}}  \bX_{\text{std}}   
\end{equation}
where
\begin{equation}
\label{eq:bstd}
\bb_{\text{std}}=\bV^{1/2} \bb \sigma^{-1}_Y = \bP^{-1} \bP_{\bX Y}
\end{equation}
are the \emph{standardized regression coefficients}.
The standardized intercept $a_{\text{std}} = 0$ vanishes because of the centering.

\subsection{Estimation of regression coefficients}

In practice, the parameters $a$ and $\bb$ are unknown. 
Therefore, to predict the response $\hat y$ for data $\bx$ using
$
\hat y =  \hat a + \hat \bb^T \bx
$
we have to learn $\hat a$ and $\hat \bb$ from some training data.
In our notation the observations
 $\bx_i$ with $i \in \{1, \ldots, n\}$ correspond to the 
random variable $\bX$, 
$y_i$ to $Y$, and $\hat y_i$ to $Y^\star$.

For estimation we distinguish between
two main scenarios.
In the large sample case with $n \gg d$
we simply replace in \eqcite{eq:acoef} and 
\eqcite{eq:bcoef} the means and covariances  by their 
\emph{empirical estimates} 
$\hat\mu_Y$, $\hat\bmu$,
$\hat \bSigma = \bS$,
$\hat \bSigma_{\bX Y} = \bS_{\bX Y}$, etc.  This gives the 
standard (and asymptotically optimal) ordinary least squares (OLS) estimates
$\hat\bb_\text{OLS} =  \bS^{-1} \,  \bS_{\bX Y}  $
and
$\hat a_\text{OLS} = \hat\mu_Y - \hat\bb^T_\text{OLS} \, \hat\bmu$.
Similarly, the coefficient of determination $R^2 = 1 - \frac{\text{RSS}}{\text{TSS}}$
is the empirical estimate of  $\Omega^2$  (cf. \tabcite{tab:vardecomp}).
If unbiased variance estimates are used the
adjusted coefficient of determination $R^2_\text{adj} = 1 - \frac{\text{RSS}/(n-d-1)}{\text{TSS}/(n-1)} $
is obtained as an alternative estimate of $\Omega^2$.
For data $\bX$ and $Y$ normally distributed it is also possible to
derive exact distributions of the estimated quantities. For example, 
the null density of the empirical squared multiple correlation coefficient
$\widehat\Omega^2 = R^2$
is $f(R^2) = \text{Beta}\left(R^2 ; \frac{d}{2}, \frac{n-d-1}{2}\right)$.

Conversely, in a ``small $n$, large $d$'' setting we use
 \emph{regularized estimates} of the covariance matrices
$\bSigma$ and $\bSigma_{\bX Y}$. For example, using 
James-Stein-type shrinkage estimation
leads to the regression approach of \citet{OS07c},
and employing  penalized maximum
likelihood inference results in scout regression \citep{WT09},
which depending on the choice
of penalty includes 
 elastic net \citep{ZH2005} and lasso \citep{Tib96} 
 as special cases.

%\newpage
\section{Measuring variable importance}

Variable importance may be defined in many different ways, see 
\citet{Fir1998} for an overview.
Here, we consider a variable to be ``important'' if it is informative
about the response and thus if its inclusion
in the predictor increases the explained variance or,
equivalently, reduces the prediction error.
To quantify the importance $\phi(X_j)$ of the explanatory 
variables $X_j$  a large number of criteria  have been suggested \citep{Groe2007}.
Desired properties of such a measure include that it 
decomposes the multiple correlation coefficient
  $\sum_{j=1}^{d} \phi(X_j) = \Omega^2$, that 
each $\phi(X_j) \geq 0$ is non-negative, and that the decomposition 
respects orthogonal subgroups
\citep{Gen1993}. The latter implies for a correlation matrix $\bP$
with block structure that the sum of the $\phi(X_j)$
of all variables $X_j$ within a block is equal to the squared multiple correlation
coefficient of that block with the response.

\subsection{Marginal correlation}
If there is \emph{no correlation among predictors} (i.e.\ if $\bP = \bI$)
 then there is general agreement
that the marginal correlations $\bP_{\bX Y} = (\rho_{1}, \ldots, \rho_{d})^T$ 
provide an optimal way to
rank variable \citep[e.g.][]{FL2008}. 
In this special case the predictor equation (\eqcite{eq:linpredstd})
simplifies to
$$
Y^\star_{\text{std}} = \bP_{\bX Y}^T \bX_{\text{std}} \,. 
$$
For $\bP = \bI$ the marginal correlations
represent the influence of each standardized covariate 
in predicting
the standardized response. Moreover, 
in this case the sum of the squared marginal correlations
$\Omega^2 = \sum_{j=1}^d \rho^2_j$
equals  the squared multiple correlation coefficient.
 Thus, the  contribution of  each variable $X_j$ 
to reducing relative prediction error is $\rho^2_j$  --- recall from \tabcite{tab:vardecomp} that
$\var(Y - Y^\star)/\sigma^2_Y=  1-  \Omega^2$. For this reason
in the uncorrelated setting 
$$
\phi^{\text{uncorr}}(X_j) = \rho^2_j
$$ 
is justifiably  the canonical
measure of variable importance for $X_j$.

However, for general $\bP$, i.e.\ in the presence of correlation among predictors,
the squared marginal correlations do not provide a decomposition of $\Omega^2$
as $\bP_{\bX Y}^T \bP_{\bX Y} \neq \Omega^2$. Thus, they are not suited
as a general variable importance criterion.

\subsection{Standardized regression coefficients}

From \eqcite{eq:linpredstd} one may consider 
standardized regression coefficients $\bb_{\text{std}}$ (\eqcite{eq:bstd})
as generalization of marginal correlations to the case of correlation
among predictors.  However, while the $\bb_{\text{std}}$
properly reduce to marginal
correlations for $\bP = \bI$ the standardized regression coefficients
also do not lead to a decomposition
of $\Omega^2$ as $\bb_{\text{std}}^T \bb_{\text{std}} =  \bP_{Y \bX }  \bP^{-2} \bP_{\bX Y}  \neq \Omega^2$.  Further objections to using $\bb_{\text{std}}$
as a measure of variable importance are discussed in \citet{Bri1994}.

\subsection{Partial correlation}

Another common way to rank predictor variables and to assign 
 $p$-values is by means
of \emph{$t$-scores} $\btau_{\bX Y} = (\tau_{1}, \ldots, \tau_{d})^T $
(which in some texts are also called standardized regression coefficients even
though they are not to be confused with $\bb_{\text{std}}$).
The $t$-scores 
are directly computed from regression coefficients via
\begin{equation*}
\begin{split}
\btau_{\bX Y} &=  \text{diag}\{\bP^{-1}\}^{-1/2} \, \bb_{\text{std}} \, (1-\Omega^2)^{-1/2} \, \sqrt{\text{d.f.}} \\
      &= \text{diag}\{\bSigma^{-1}\}^{-1/2}      \, \bb \, \sigma^{-1}_Y (1-\Omega^2)^{-1/2} \, \sqrt{\text{d.f.}} \, .
\end{split}
\end{equation*}
The constant $\text{d.f.}$ is the degree of freedom and $\text{diag}\{\bM\}$
the matrix $\bM$ with its off-diagonal entries set to zero. 

Completely equivalent to $t$-scores in terms of variable ranking
are the \emph{partial correlations} $\tilde\bP_{\bX Y} = (\tilde\rho_{1}, \ldots, \tilde\rho_{d})^T$ between 
the response $Y$ and predictor $X_j$ conditioned on all the remaining predictors $X_{\neq j}$.
The $t$-scores can be converted to partial correlations 
using the relationship
$$
\tilde\rho_{j}  = \tau_{j} / \sqrt{ \tau^2_{j} + \text{d.f.} } \, \,.
$$
Interestingly, the value
of $\text{d.f.}$  specified in the $t$-scores cancels out when computing $\tilde\rho_{j}$.
An alternative but equivalent route to obtain the partial correlations
is by inversion and subsequent standardization of the
joined correlation matrix of $Y$ and $\bX$ \citep[e.g.][]{OS07c}.

The $p$-values computed in many statistical
software packages for each variable in a linear model are 
based on empirical estimates of $\btau_{\bX Y}$ with $\text{d.f.}=n-d-1$.
Assuming normal $\bX$ and $Y$ the null distribution of $\hat\tau_{j}$  is Student  
$t$
with $n-d-1$ degrees of freedom.  Exactly the same $p$-values are obtained
from the empirical
partial correlations $\tilde{r}_{j}$ which have null-density 
$f(\tilde{r}_{j}) = |\tilde{r}_{j}| \, \text{Beta}\left(\tilde{r}_{j}^2 ; \frac{1}{2}, \frac{\kappa-1}{2}\right)$
with $\kappa=\text{d.f.}+1 = n-d$ and $\var(\tilde{r}_{j}) = \frac{1}{\kappa}$.

Despite being widely used, a key problem of partial correlations $\tilde\bP_{\bX Y}$ (and hence also of the
corresponding $t$-scores) 
for use in variable ranking and assigning variable importance is
that in the case of vanishing correlation $\bP = \bI$  they do \emph{not} 
properly reduce
to the marginal correlations $\bP_{\bX Y}$.    This can be seen already from the simple
case with three variables  $Y$, $X_1$, and $X_2$
with partial correlation 
$$
\rho_{Y,X_1|X_2} = {\rho_{Y,X_1} - \rho_{Y,X_2} \rho_{X_1,X_2} \over \sqrt{1-\rho_{Y,X_2}^2} \sqrt{1-\rho_{X_1,X_2}^2}}
$$ 
which for $\rho_{X_1,X_2} = 0$ is not identical to $\rho_{Y,X_1}$ unless
$\rho_{Y,X_2}$ also vanishes.

\subsection{Hoffman-Pratt product measure}

First suggested by \citet{Hof1960} and later defended by \citet{Pra1987}
is the following alternative measure of
variable importance
$$
\phi^{\text{HP}} (X_j) =  (\bb_{\text{std}})_j \,\rho_j =  (\bP^{-1} \bP_{\bX Y})_j \, \rho_j \,.
$$
By construction, $\sum_{j=1}^d \phi^{\text{HP}} (X_j) = \Omega^2$, 
and if correlation among predictors is zero then $\phi^{\text{HP}} (X_j) = \rho_j^2$.
Moreover, the Hoffman-Pratt measure satisfies the 
orthogonal compatibility
criterion \citep{Gen1993}. 

However, in addition to these desirable properties 
the Hoffman-Pratt variable importance measure also exhibits two severe defects. First, $\phi^{\text{HP}} (X_j)$ 
may become negative, and second the relationship of the Hoffman-Pratt
measure with the original predictor equation is unclear.
Therefore, the use of $\phi^{\text{HP}} (X_j)$ is discouraged
by most authors \citep[cf.][]{Groe2007}.

\subsection{Genizi's measure}

More recently, \citet{Gen1993} proposed the variable importance measure
$$
\phi^{\text{G}} (X_j) =  \sum_{k=1}^d \left( (\bP^{1/2})_{jk}  \,  (\bP^{-1/2}  \bP_{\bX Y} )_k \right)^2   \,.
$$
Here and in the following $\bP^{1/2}$ is the uniquely defined 
matrix square root with $\bP^{1/2}$ symmetric and positive definite.

Genizi's measure provides a decomposition
$\sum_{j=1}^d \phi^{\text{G}} (X_j) = \Omega^2$, 
reduces to the squared marginal correlations in case of no
correlation, and obeys the orthogonality criterion.  In contrast to 
$\phi^\text{HP}(X_j)$ the Genizi measure is by construction also  non-negative, 
$\phi^{\text{G}} (X_j) \geq 0$. 

However, like the Hoffman-Pratt measure the connection of $\phi^{\text{G}} (X_j)$
with the original predictor equations is unclear.

%\newpage
\section{Variable selection using CAR scores}

In this section we introduce 
CAR scores $\bomega = (\omega_1, \ldots, \omega_d)^T$
and the associated variable importance measure $\phi^{\text{CAR}}(X_j) = \omega_j^2$
and discuss their use in variable selection.

Specifically, we argue that CAR scores $\bomega$  and $\phi^{\text{CAR}}(X_j)$
naturally generalize  marginal correlations
$\bP_{\bX Y} = (\rho_1, \ldots, \rho_d)^T$ and the importance measure
$\phi^{\text{uncorr}}(X_j) = \rho_j^2$ to settings with
non-vanishing correlation $\bP$
among explanatory variables.

\subsection{Definition of the CAR score}

The CAR scores $\bomega$ are defined as
\begin{equation}
\label{eq:carscore}
\bomega  = \bP^{-1/2} \, \bP_{\bX Y} \, ,
\end{equation}
i.e.\ as the marginal correlations $\bP_{\bX Y}$ adjusted by the factor $\bP^{-1/2}$. 
Accordingly, the acronym ``CAR'' is an 
abbreviation for 
Correlation-Adjusted (marginal) coRrelation.
The CAR scores $\bomega$ are constant population quantities and
 not random variables.  

\begin{table}[!t]
\caption{Relationship between CAR scores 
$\bomega$ and common quantities from the linear model.}
\begin{center}
\begin{tabular}{llcl}
\toprule
Criterion & \multicolumn{3}{c}{Relationship with CAR scores $\bomega$}\\
\midrule
\midrule
Regression coefficient & $\bb = \bSigma^{-1/2}  \bomega \,  \sigma_Y$  & $\leftrightarrow$ & $ \bomega = \bSigma^{1/2}  \bb \,  \sigma_Y^{-1}$ \\
Standardized regression coeff. & $\bb_{\text{std}} = \bP^{-1/2}  \bomega$  & $\leftrightarrow$& $ \bomega = \bP^{1/2}  \bb_{\text{std}}$ \\
Marginal correlation & $\bP_{\bX Y} = \bP^{1/2}  \bomega$  &$\leftrightarrow$ & $ \bomega = \bP^{-1/2} \bP_{\bX Y} $ \\
Regression $t$-score &  \multicolumn{3}{l}{ $\btau_{\bX Y} =  (\bP\, \text{diag}\{\bP^{-1}\})^{-1/2} \, \bomega \, (1-\bomega^T \bomega)^{-1/2} \sqrt{\text{d.f.}}$ } \\

\bottomrule 
\end{tabular}
\end{center}
\label{tab:relationships}
\end{table}

\tabcite{tab:relationships} summarizes some connections of
CAR scores with various other quantities from the linear model.
For instance, CAR scores may be viewed as intermediates 
between marginal correlations and standardized regression coefficients.
If correlation among predictors vanishes 
 the CAR scores become identical to the 
marginal correlations.

The CAR score is a relative of the CAT score (i.e.\ correlation-adjusted $t$-score)
that we have introduced previously as variable ranking statistic for classification
problems \citep{ZS09}.  In \tabcite{tab:catcar} we review some properties of the CAT score in
comparison with the CAR score.  In particular, in the CAR score the 
marginal correlations $\bP_{\bX Y}$
play the same role as the  $t$-scores $\btau$ in the CAT score.

\subsection{Estimation of CAR scores}

In order to obtain estimates  $\hat\bomega$ of the CAR scores we
substitute in \eqcite{eq:carscore} suitable estimates of the two 
matrices $\bP^{-1/2}$ and $\bP_{\bX Y}$.  For large sample sizes $n \gg d$ we 
suggest using empirical  and for small sample size shrinkage
estimators, e.g. as in \citet{SS05c}. An efficient algorithm for calculating
the inverse matrix square-root $\bR^{-1/2}$ 
for the shrinkage correlation estimator
is described in \citet{ZS09}.  If the correlation matrix exhibits
a known pattern, e.g., a block-diagonal structure, then it is advantageous
to employ a correspondingly structured estimator.

The null distribution of 
the \emph{empirical} CAR scores under normality is
identical to that of the empirical marginal correlations. Therefore,
regardless of the value of $\bP$  
the null-density is 
$f(\hat\omega_j) = | \hat\omega_{j} | \text{Beta}\left(\hat\omega_{j}^2 ; \frac{1}{2}, \frac{\kappa-1}{2}\right)$
with $\kappa=n-1$.   

\begin{table}[!t]
\caption{Comparison of CAT and CAR scores.}
\begin{center}

\begin{tabular}{ l c c  }
\toprule
    & CAT & CAR \\
\midrule
 Response $Y$ & Binary & Metric \\
 Definition & $\btau^{\text{adj}}=\bP^{-1/2} \btau $ & $\bomega=\bP^{-1/2} \bP_{\bX Y} $ \\
 Marginal quantity & $\btau = (\frac{1}{n_1}+\frac{1}{n_2})^{-1/2} \bV^{-1/2}(\bmu_1-\bmu_2)$ & $ \bP_{\bX Y} $\\
 Decomposition & Hotelling's $T^2$    & Squared multiple correlation\\
    & $ T^2= \sum_{j=1}^d (\tau^{\text{adj}}_j)^2$ &  $ \Omega^2 = \sum_{j=1}^d \omega_j^2$ \\
 Global test statistic \\
\hspace{2mm} for a set of size $s$   &    $T^2_s = \sum_{j=1}^s (t^{\text{adj}}_j)^2 $ &   $R^2_s = \sum_{j=1}^s \hat\omega_j^2$\\ 
Null distribution for \\
\hspace{2mm} empirical statistic & $T^2_s (\frac{m-s+1}{m s}) \sim F(s, m-s+1)$  & $R^2_s \sim \text{Beta}(\frac{s}{2}, \frac{n-s-1}{2})$ \\
\hspace{2mm} under normality     & with $m=n_1+n_2-2$ \\
\bottomrule 
\end{tabular}

\end{center}
\label{tab:catcar}
\end{table}

\subsection{Best predictor in terms of CAR scores}

Using CAR scores  the best linear predictor
 (\eqcite{eq:linpredstd})
can be written in the simple form
\begin{equation}
\label{eq:linpredcar}
Y^\star_{\text{std}}  = \bomega^T \bdelta(\bX) = \sum_{j=1}^d \omega_j \delta_j(\bX) \,  ,
\end{equation}
where
\begin{equation}
\label{eq:mahalanobis}
\bdelta(\bX) = \bP^{-1/2}  \bV^{-1/2} ( \bX - \bmu ) = \bP^{-1/2} \bX_{\text{std}} \, 
\end{equation}
are the Mahalanobis-decorrelated and standardized predictors
with $\var(\bdelta(\bX)) = \bI$.
Thus, the CAR scores $\bomega$ are the weights that
describe the influence of each decorrelated variable
in predicting the standardized response.
Furthermore, 
with $\corr(\bX_{\text{std}}, Y ) = \bP_{\bX Y}$
we have 
$$
\bomega = \corr(\bdelta(\bX), Y ) \, ,
$$
i.e.\ CAR scores are the correlations
between the response and the decorrelated covariates.

\subsection{Special properties of the Mahalanobis transform}

The computation of CAR score relies on decorrelation of
predictors using \eqcite{eq:mahalanobis} which is known as
the Mahalanobis transform.  
Importantly, the Mahalanobis transform has a number of  properties
not shared by other decorrelation transforms  with $\var(\bdelta(\bX)) = \bI$.
First, it is the unique linear transformation that
minimizes
$\expect\left((\bdelta(\bX)-\bX_{\text{std}})^T (\bdelta(\bX)-\bX_{\text{std}})\right)$,
see \citet{Gen1993} and \citet[][Section 6.5]{HKO2001}.  Therefore,
the Mahalanobis-decorrelated predictors $\bdelta(\bX)$
 are nearest to the original standardized predictors $\bX_{\text{std}}$.
Second, as $\bP^{-1/2}$ is positive definite 
$\bdelta(\bX)^T \bX_{\text{std}} > 0$ for any $\bX_{\text{std}}$
which implies that the decorrelated  and the standardized predictors
are informative about each other also on a componentwise level
(for example they must have the same sign).  The correlation of the corresponding
elements in  $\bX_{\text{std}}$ and $\bdelta(\bX)$ is given by
$\corr( (\bX_{\text{std}})_i, \bdelta(\bX)_i = (\bP^{1/2})_{ii}$.

\subsection{Comparison of CAR scores and partial correlation}

Further insights into the interpretation of CAR scores can be gained by 
a comparison with partial correlation.

The partial correlation between $Y$ and a predictor $X_i$ is obtained
by first removing the linear effect of the remaining $d-1$ predictors $X_{\neq i}$
from both $Y$ and $X_i$ and subsequently computing the correlation between
the  respective remaining residuals.

In contrast, with CAR scores the response $Y$ is left unchanged whereas
all $d$ predictors are simultaneously orthogonalized, i.e.\ the linear effect
of the other variables $X_{\neq i}$ on $X_i$ is  removed simultaneously from all predictors \citep[][Section 6.5]{HKO2001}. Subsequently, the CAR score is found
as the correlation between the ``residuals'', i.e.\ the unchanged response and the decorrelated predictors.
Thus, CAR scores may be viewed as a multivariate variant of the so-called  part correlations.

\subsection{Variable importance and error decomposition}

The squared multiple correlation coefficient is the sum of the
squared CAR scores,
$\Omega^2 = \bomega^T \bomega = \sum_{j=1}^d \omega_j^2$.
Consequently, the
 nominal mean squared prediction error in terms of CAR scores can be written
$$
\expect((Y - Y^\star)^2) = \sigma^2_Y \, (1-  \bomega^T \bomega) \,, 
$$
which implies that (decorrelated) variables with small CAR scores contribute little to
improve the prediction error or to
reduce the  unexplained variance.
This suggests to define
$$
\phi^\text{CAR}(X_j) = \omega_j^2 
$$
as a measure of variable importance.
$\phi^\text{CAR}(X_j)$ is always non-negative, reduces to $\rho_j^2$ for uncorrelated explanatory variables, and  leads to the canonical decomposition 
$$
\Omega^2 = \sum_{j=1}^d \phi^\text{CAR}(X_j) \,.
$$
Furthermore, it is easy to see that  $\phi^\text{CAR}(X_j)$
 satisfies the orthogonal compatibility
criterion demanded in \citet{Gen1993}.
Interestingly, Genezi's own importance measure $\phi^{\text{G}} (X_j)$
can be understood as a weighted average 
$
\phi^{\text{G}} (X_j) =  \sum_{k=1}^d (\bP^{1/2})^2_{jk} \, \phi^\text{CAR}(X_k)      
$
of squared CAR scores. 

In short, what we propose here is to first Mahalanobis-decorrelate the predictors
to establish a canonical basis, and subsequently we define the  importance of a variable $X_j$
as the natural weight $\omega_j^2$ in this reference frame.

\subsection{Grouped CAR score}

Due to the additivity  of squared car scores it is straightforward
to define a \emph{grouped CAR score}  for a set of variables
as the sum of the individual squared CAR scores
$$
\omega_{\text{grouped}} =   
 \sqrt{ \sum_{g \in \text{set}} \omega^2_g } \,.
$$
As with the grouped CAT score \citep{ZS09}  we also may add a sign in this
definition.

An estimate of the squared grouped CAR score is an example of
a simple global test statistic that may be useful, e.g., in studying gene set enrichment \citep[e.g.][]{AS09}.
The null density of the empirical estimate $R^2_s = \sum_{j=1}^s \hat\omega^2_j$ for a set of size $s$
is given by $f(R^2_s) = \text{Beta}(R^2_s; \frac{s}{2}, \frac{n-s-1}{2})$ which for
$s=1$ reduces to the null distribution of the squared empirical  CAR score, and for $s=d$ equals
the distribution of the squared empirical multiple correlation coefficient $R^2$. 

Another related summary (used in particular in the next section) is the accumulated squared CAR score 
$\Omega^2_k$ for
the largest $k$ predictors.
Arranging
the CAR scores in decreasing order of absolute magnitude
$\omega_{(1)}, \ldots, \omega_{(d)}$ with $\omega_{(1)}^2 > \ldots > \omega_{(d)}^2$
this can be written as
$$
\Omega^2_k = \sum_{j=1}^k \omega_{(j)}^2 \,.
$$

\subsection{CAR scores and information criteria for model selection}

CAR scores define a canonical ordering of the explanatory variables.
Thus, variable selection using CAR scores is a simple matter of
thresholding (squared) CAR scores.  Intriguingly, this provides
a direct link to model selection procedures
using information criteria such as AIC or BIC. 

Classical model selection can be put into the 
framework of penalized residual sum of squares \citep{Geo2000}
with
$$
\text{RSS}_k^\text{penalized} = \text{RSS}_k + \lambda k \, \hat\sigma^2_\text{full}  \, ,
$$
where $k \leq d$ is the number of included predictors and
$\hat\sigma^2_\text{full}$
an estimate of the variance of the residuals using the full model
with all predictors included. The model selected as optimal minimizes
 $\text{RSS}_k^\text{penalized}$, with the penalty parameter $\lambda$
fixed in advance.  
The choice of $\lambda$ corresponds to the choice of information
criterion --- see
\tabcite{tab:modsel} for details.

With 
$\text{RSS}_k /(n \hat\sigma^2_Y)$ as empirical estimator of $1 -\Omega^2_k$,
and $R^2$ as estimate of $\Omega^2$,
we rewrite the above as
\begin{equation*}
\begin{split}
\frac{\text{RSS}_k^\text{penalized}}{n \hat\sigma^2_Y} &=  
1 - \widehat\Omega^2_k + \frac{\lambda k (1-R^2)}{n} \\ 
&= 1 - \sum_{j=1}^k \biggl(\hat\omega_{(j)}^2 - \frac{\lambda (1-R^2)}{n}\biggr) \, .
\end{split}
\end{equation*}
This quantity decreases with $k$ as long as $\hat\omega_{(k)}^2 > \hat\omega_c^2 = \frac{\lambda (1-R^2)}{n} $.
Therefore, in terms of CAR scores
 classical model selection is equivalent to thresholding $\hat\omega_{j}^2$
at critical level $\hat\omega_c^2$, where
predictors with  $\hat\omega_{j}^2 \leq \hat\omega_c^2$ are removed.
If $n$ is large or for a perfect fit ($R^2 = 1$)
 all predictors
are retained.

\begin{table}[t]
\caption{Threshold parameter $\lambda$ for some classical model selection procedures.}
\begin{center}
\begin{tabular}{lll}
\toprule
Criterion & Reference & Penalty parameter \\
\midrule
\midrule
AIC & \citet{Aka74} & $\lambda=2$ \\
$C_p$ & \citet{Mal1973} & $\lambda=2$ \\
BIC & \citet{Schwarz1978} & $\lambda=\log(n)$ \\
RIC & \citet{FG1994} & $\lambda = 2 \log(d)$ \\
\bottomrule 
\end{tabular}
\end{center}
\label{tab:modsel}
\end{table}

As alternative to using a fixed cutoff we may also conduct model selection
with an adaptive choice of threshold.
One such  approach is to remove null-variables by controlling 
false non-discovery rates (FNDR) as described in \citet{AS2010}.
The required null-model for computing FNDR from observed CAR scores $\hat\omega_j$ is the same
as when using marginal correlations.
Alternatively, an optimal threshold may be chosen, e.g., by minimizing
cross-validation
estimates of  prediction error.

\subsection{Grouping property, antagonistic variables and oracle CAR score}

A favorable feature of the elastic net procedure for
variable selection is the ``grouping property'' which
 enforces the simultaneous selection 
of highly correlated predictors \citep{ZH2005}.
Model selection using CAR scores
also exhibits the grouping property because
predictors that are highly correlated have nearly 
identical CAR scores.  This can directly be seen from 
the definition $\bomega = \bP^{1/2}  \bb_{\text{std}}$ of the CAR score.  
For two predictors $X_1$ and $X_2$
and correlation $\corr(X_1, X_2) = \rho$ 
a simple algebraic calculation shows that the difference between the 
two squared CAR scores equals
$$
\omega_1^2-\omega_2^2 = \left(\, (\bb_{\text{std}})_1^2 -(\bb_{\text{std}})_2^2 \,\right) \sqrt{1 - \rho^2} \,.
$$
Therefore, the two squared CAR scores become identical with growing 
absolute value of the correlation between the variables. This grouping 
property is intrinsic to the CAR score itself 
and not a property of an estimator. 

In addition to the grouping property the CAR score also exhibits an important
behavior with regard to antagonistic variables.  If the regression coefficients
of two variables have opposing signs and these variables are in addition positively correlated then the corresponding CAR scores decrease to zero. For example, with
$(\bb_{\text{std}})_2 = -(\bb_{\text{std}})_1$ we get
$$
\omega_1= -\omega_2 = (\bb_{\text{std}})_1 \sqrt{1-\rho} \, . 
$$
This implies that antagonistic positively correlated variables will be bottom ranked.
 A similar effect occurs for  protagonistic
variables that are negatively correlated, as with $(\bb_{\text{std}})_1 = (\bb_{\text{std}})_2$
we have 
$$
\omega_1= \omega_2 = (\bb_{\text{std}})_1 \sqrt{1+\rho} \, ,
$$
which decreases to zero for large negative correlation (i.e.\ for $r \rightarrow -1$).

Further insight into the CAR score is obtained by 
considering an ``oracle version''
where it is known in advance which predictors are truly non-null.
Specifically, we assume that the regression coefficients can
be written as
$$
\bb_{\text{std}} = \left( \begin{array}{c}
\bb_\text{std, non-null}  \\
0 \end{array} \right) 
$$
and that there is no correlation between null and non-null variables
so that
the correlation matrix $\bP$ has block-diagonal structure
$$
\bP = \left( \begin{array}{cc}
\bP_\text{non-null} & 0 \\
0 & \bP_\text{null} \end{array} \right) \,.
$$
The resulting oracle CAR score 
$$
\bomega = \bP^{1/2}  \bb_{\text{std}} = \left( \begin{array}{c}
\bomega_\text{non-null}  \\
0 \end{array} \right) 
$$
is exactly zero for the null variables.  Therefore, asymptotically
the null predictors will be identified by the CAR score
with probability one as long as the employed estimator is consistent.

%\newpage
\section{Applications}

In this section we demonstrate variable selection by thresholding CAR scores
in a simulation study and by analyzing experimental data.
As detailed below, we considered  large and small sample settings
for both synthetic and real data.  

\subsection{Software}

All analyzes were done using
the R platform \citep{RPROJECT}.  A corresponding R package ``care''
implementing CAR estimation and CAR regression 
is available from the authors' web page (\url{http://www.strimmerlab.org/software/care/}) 
and also from the CRAN archive (\url{http://cran.r-project.org/web/packages/care/}).  The code for the computer simulation is also available from
our website.

For comparison we fitted in our study 
lasso and elastic net regression models using 
the 
algorithms available in the R package ``scout'' \citep{WT09}.
In addition, we employed the boosting algorithm for linear models
as   implemented in the R package
``mboost''  \citep{HB2006}, ordinary least squares with no
variable selection (OLS), with partial correlation ranking (PCOR)
and with variable ranking by the Genizi method.

\subsection{Simulation study}

In our simulations we broadly followed the setup 
employed in \citet{ZH2005}, \citet{WT09} and \citet{WN+2011}.

Specifically, we considered the following scenarios:
\begin{itemize}
\item \emph{Example 1:} 8 variables with $\bb = (3, 1.5, 0, 0, 2, 0, 0, 0)^T$.  
                 The predictors exhibit autoregressive
                 correlation with $\corr(X_j, X_k) = 0.5^{|j-k|}$.
\item \emph{Example 2:} As Example 1 but with $\corr(X_j, X_k) = 0.85^{|j-k|}$.
\item \emph{Example 3:} 40 variables with $\bb = (3, 3, 3, 3, 3, -2, -2, -2, -2, -2, 0, \ldots, 0)^T$.
                 The correlation between all pairs of the first 10 variables is set to 0.9, 
                 and otherwise set to 0.
\item \emph{Example 4:} 40 variables with $\bb = (3, 3, -2, 3, 3, -2, 0, \ldots, 0)^T$.
                 The pairwise correlations among the first three variables 
                 and among the second three variables equals 0.9 and is otherwise
                 set to 0.
\end{itemize}
The intercept was set to $a=0$ in all scenarios.
We generated samples $\bx_i$ by drawing from a multivariate normal distribution 
with unit variances, zero means and correlation structure $\bP$ as indicated for
each simulation scenario.
To compute $y_i = \bb^T \bx_i + \varepsilon_i$ we sampled the error $\varepsilon_i$
from a normal distribution with zero mean and standard deviation $\sigma$
(so that $\var(\varepsilon) = \var(Y-Y^\star) = \sigma^2)$. 
In Examples~1 and~2 the dimension is $d=8$ and the sample sizes considered
were $n=50$ and $n=100$ to represent a large sample setting.  In contrast,
for Examples 3 and 4 the dimension is $d=40$
and sample sizes were small (from $n=10$ to $n=100$).
In order to vary the ratio of signal and noise variances we used different 
degrees of unexplained variance ($\sigma=1$ to  $\sigma=6$).
For fitting the regression models we employed a training data set of 
size $n$.   The tuning parameter of each approach was optimized using
an additional independent validation 
data set of the same size $n$.  In the CAR, PCOR and Genizi approach the tuning parameter corresponds directly
to the number of included variables, whereas for elastic net, lasso, and boosting
the tuning parameter(s) corresponds to a regularization parameter.

For each estimated set of regression coefficients
$\hat\bb$
we computed the model error and the model size.
All simulations were repeated 200 times,
and the average relative model error as well as the median model size
was reported.   For estimating CAR scores and associated regression coefficients
we used in the large sample cases (Examples~1 and 2)
 the empirical estimator and
and otherwise (Examples~3 and 4) shrinkage estimates.

\subsection{Results from the simulation study}

\begin{table}[!p]
\caption{Average relative model error (x 1000) and its standard deviation
as well as the mean true and false positives (TP+FP) 
in alternating rows for Examples 1 and 2.   These simulations 
represent large sample settings ($d=8$ with $n=40$ to $n=100$).}
\begin{center}
\small
\begin{tabular}{lrrrrrrr}
\toprule
            &  CAR ${}^*$    & Elastic Net & Lasso  & Boost   & OLS  & PCOR & Genizi \\ 
\midrule
\midrule
\multicolumn{7}{l}{Example 1 (true model size = 3)}  \\
$n=50$ \\
$\sigma=1$  & \bf 107 (5)  & 135 (7)  & 132 (6)  & 390 (24)  & 217 (8) & \bf 107 (5) & 109 (6) \\
            & 3.0+1.2      & 3.0+1.9  & 3.0+1.8  & 3.0+2.6   & 3.0+5.0 & 3.0+0.7  & 3.0+1.3 \\
$\sigma=3$  & \bf 119 (7)  & 130 (6)  & 148 (6)  & 151 (6)   & 230 (9) & 153 (8)  & 129 (7) \\
            & 3.0+1.3      & 3.0+2.6  & 3.0+1.9  & 3.0+3.5   & 3.0+5.0 & 2.9+0.9  & 3.0+1.3 \\
$\sigma=6$  & 143 (6)   & \bf 127 (5) & 152 (6)  & 149 (8)   & 227 (8) & 163 (6)  & 139 (6) \\
            & 2.5+1.2      & 2.8+2.4  & 2.6+2.0  & 2.8+3.7   & 3.0+5.0 & 2.3+1.4  & 2.5+1.1 \\
$n=100$ \\
$\sigma=1$  & \bf 53 (3)   & 64 (3)   & 59 (3)   & 219 (18)  & 97 (4)  & 54 (3)   & 55 (3)  \\
            & 3.0+1.0      & 3.0+1.9  & 3.0+1.5  & 3.0+2.4   & 3.0+5.0 & 3.0+0.8  & 3.0+1.2 \\
$\sigma=3$  & \bf 55 (3)   & 58 (2)   & 59 (3)   & 78 (3)    & 99 (3)  & 59 (3)   & 56 (4)  \\
            & 3.0+1.2      & 3.0+2.1  & 3.0+1.9  & 3.0+3.6   & 3.0+5.0 & 3.0+0.8  & 3.0+1.0 \\
$\sigma=6$  & 65 (3)     & \bf 64 (3) & 69 (3)   & 66 (3)    & 97 (3)  & 76 (3)   & 65 (3)  \\
            & 2.8+1.2      & 2.9+2.4  & 2.9+2.1  & 3.0+3.7   & 3.0+5.0 & 2.6+1.3  & 2.8+1.5 \\
\multicolumn{7}{l}{Example 2  (true model size = 3)} \\
$n=50$ \\
$\sigma=1$  & \bf 110 (5)  & 147 (7)  & 134 (6)  & 716 (55)  & 230 (9) & 120 (8)  & 130 (6) \\
            & 3.0+1.4      & 3.0+2.4  & 3.0+2.0  & 3.0+3.1   & 3.0+5.0 & 3.0+0.9  & 3.0+2.3 \\
$\sigma=3$  & 127 (5)  & \bf 124 (5)  & 139 (6)  & 165 (7)   & 220 (8) & 178 (9)  & 158 (8) \\
            & 2.8+1.6      & 3.0+3.0  & 2.8+2.2  & 2.8+3.5   & 3.0+5.0 & 2.4+1.6  & 2.8+2.1 \\
$\sigma=6$  & 121 (5)  & \bf 95 (4)   & 121 (6)  & 110 (5)   & 232 (9) & 165 (7)  & 135 (5) \\
            & 2.2+1.5      & 2.7+3.2  & 2.2+1.9  & 2.5+3.4   & 3.0+5.0 & 1.8+1.5  & 2.2+1.6 \\
$n=100$ \\
$\sigma=1$  & \bf 49 (3)   & 67 (3)   & 61 (3)   & 325 (28)  & 95 (3)  & 52 (3)   & 60 (3)  \\
            & 3.0+1.1      & 3.0+2.2  & 3.0+1.9  & 3.0+3.0   & 3.0+5.0 & 3.0+1.0  & 3.0+2.0 \\
$\sigma=3$  & \bf 62 (3)   & 63 (3)   & 64 (3)   & 83 (4)    & 101 (4) & 78 (4)   & 62 (4)  \\
            & 3.0+1.5      & 3.0+2.7  & 3.0+2.2  & 3.0+3.3   & 3.0+5.0 & 2.8+1.2  & 3.0+1.9 \\
$\sigma=6$  & 64 (3)   & \bf 53 (2)   & 59 (2)   & 54 (2)    & 100 (4) & 77 (3)   & 66 (3)  \\
            & 2.6+1.7      & 2.9+3.1  & 2.6+2.1  & 2.7+3.3   & 3.0+5.0 & 2.0+1.4  & 2.7+1.8 \\
\bottomrule 
\multicolumn{7}{c}{${}^*$ using empirical CAR estimator.}\\
\end{tabular}
\end{center}
\label{tab:simulations1}
\end{table}

\begin{table}[!p]
\caption{Average relative model error (x 1000) and its standard deviation
as well as the mean true and false positives (TP+FP) 
in alternating rows for Examples 3 and 4.  
These simulations
represent small sample settings ($d=40$ with $n=10$ to $n=100$).}
\begin{center}
\small
\begin{tabular}{lrrrrrrr}
\toprule
            & CAR ${}^*$ & Elastic Net & Lasso     & Boost & OLS  & PCOR & Genizi   \\ 
\midrule
\midrule
\multicolumn{7}{l}{Example 3 (true model size = 10)}  \\
$n=10$ \\
$\sigma=3$  & \bf 1482 (44) & 1501 (45) & 1905 (75) & 2203 (66) & ---        \\
            & 6.1+7.0       & 6.3+11.5  & 2.1+4.7   & 2.4+13.7  & ---        \\
$n=20$ \\
$\sigma=3$  & \bf 838 (30)  & 950 (26)  & 1041 (29) & 1421 (44) & ---        \\
            & 6.4+2.7       & 5.6+6.2   & 2.5+4.2   & 2.8+12.0  & ---        \\
$n=50$ \\
$\sigma=3$  & \bf 358 (11)  & 571 (10)  & 608 (8)   & 805 (12)  & 5032 (214) & 888 (27) & 364 (12) \\
            & 8.5+0.6       & 5.2+2.9   & 3.3+3.3   & 4.2+13.0  & 10.0+30.0  & 2.5+2.2  & 8.4+1.1  \\
$n=100$ \\
$\sigma=3$  & 172 (6)       & 488 (4)   & 525 (6)   & 569 (8)   & 693 (14)   & 406 (10) & \bf 155 (5) \\
            & 9.5+0.7       & 6.0+6.8   & 5.9+10.8  & 7.1+17.3  & 10.0+30.0  & 6.9+3.1  & 9.6+0.6 \\
\multicolumn{7}{l}{Example 4 (true model size = 6)}  \\
$n=10$ \\
$\sigma=6$  & \bf 835 (24)  & 1061 (34) & 1684 (60) & 1113 (39) & ---        \\
            & 3.5+9.3       & 4.5+20.2  & 1.6+6.4   & 1.5+9.8   & ---        \\
$n=20$ \\
$\sigma=6$  & \bf 527 (18)  & 767 (25)  & 925 (40)  & 791 (22)  & ---        \\
            & 4.2+7.0       & 4.4+13.2  & 2.4+7.5   & 2.0+9.4   & ---        \\
$n=50$ \\
$\sigma=6$  & \bf 200 (11)  & 226 (9)   & 293 (14)  & 359 (11)  & 4991 (176) & 1075 (67) & 204 (7) \\
            & 4.9+3.0       & 4.3+4.7   & 3.0+4.0   & 3.3+12.9  & 6.0+36.0   & 2.8+5.0   & 5.5+0.8 \\
$n=100$ \\
$\sigma=6$  & \bf 87 (4)    & 107 (4)   & 112 (3)   & 168 (4)   & 699 (16)   & 232 (8)   & 94 (4) \\
            & 5.4+1.2       & 4.5+2.9   & 3.5+2.8   & 3.8+12.2  & 6.0+36.0   & 4.6+1.7   & 5.8+0.9 \\
\bottomrule
\multicolumn{7}{l}{${}^*$ using shrinkage CAR estimator.}\\
\end{tabular}
\end{center}

\label{tab:simulations2}
\end{table}

%\afterpage{\clearpage}

\begin{figure}[!p]
\begin{center}
\centerline{\includegraphics[scale=0.68]{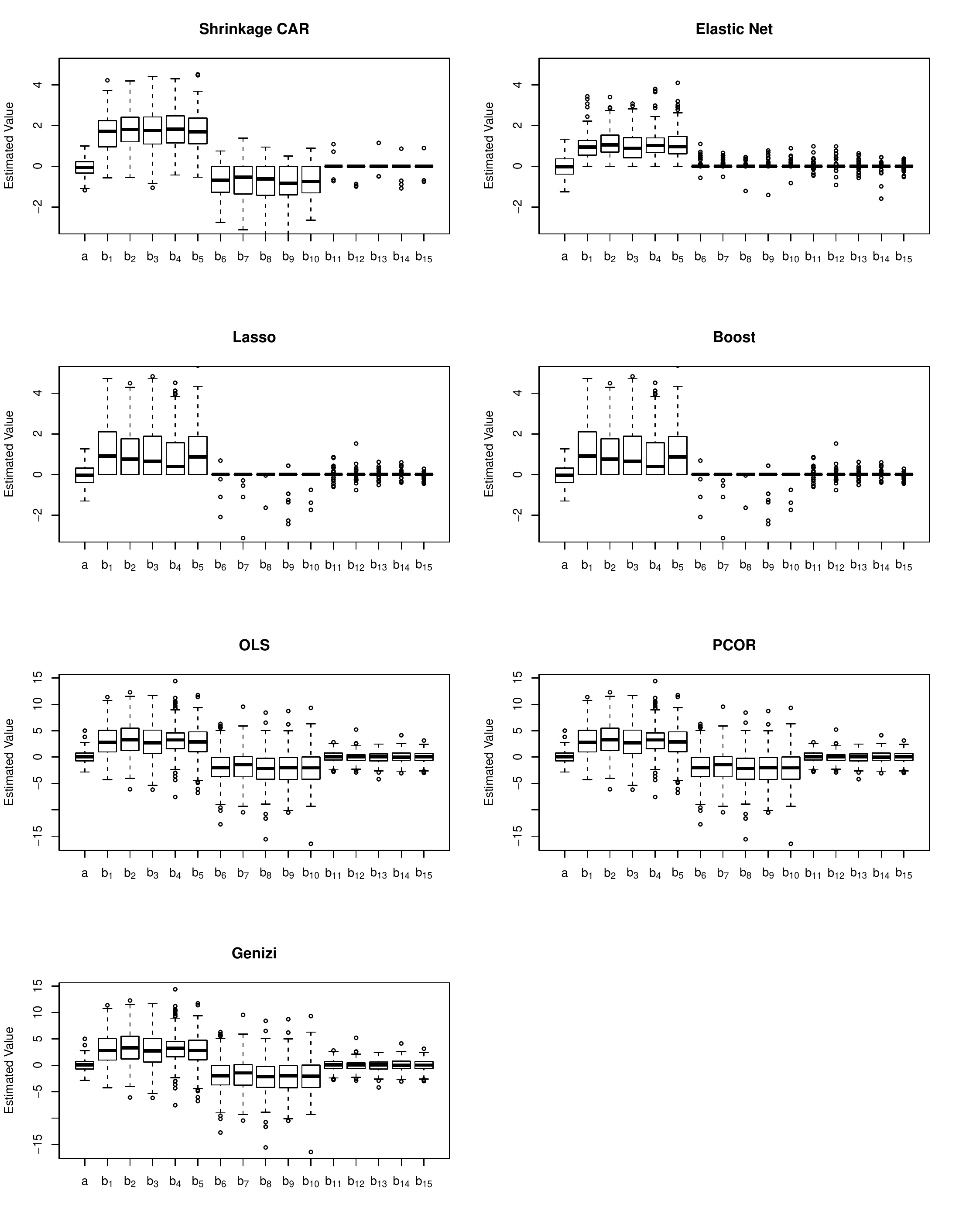}}
\caption{Distribution of estimated regression coefficients 
for Example 3 with $n=50$ and $\sigma=3$. Coefficients
for variables $X_{16}$
to $X_{40}$ are not shown but are similar to those of $X_{11}$ to $X_{15}$.
The scale of the plots for OLS, PCOR and Genizi is different from that of
the other four methods.}
\label{fig:m3beta}
\end{center}
\end{figure}

\begin{table}[t]
\caption{Population quantities for Example 1 with $\sigma=3$. 
}
\begin{center}
\begin{tabular}{lrrrrrrrr}
\toprule
Quantity           & $X_1$ & $X_2$ & $X_3$ & $X_4$ & $X_5$ & $X_6$ & $X_7$ & $X_8$  \\ 
\midrule
\midrule
$\bb$              & 3     & 1.5   & 0     & 0     & 2     & 0     & 0     &  0 \\
$\bb_{\text{std}}$ & 0.55  & 0.27  & 0     & 0     & 0.36  & 0     & 0     &  0 \\
$\tilde\bP_{\bX Y}$& 0.65  & 0.36  & 0     & 0     & 0.46  & 0     & 0     &  0 \\ 
$\bP_{\bX Y}$      & 0.70  & 0.59  & 0.36  & 0.32  & 0.43  & 0.22  & 0.11  & 0.05 \\
$\bomega$          & 0.60  & 0.40  & 0.15  & 0.13  & 0.36  & 0.10  & 0.04  & 0.02 \\
$\phi^{\text{CAR}}$& 0.36  & 0.16  & 0.02  & 0.02  & 0.13  & 0.01  & 0.00  & 0.00 \\
\bottomrule 
\multicolumn{9}{c}{Numbers are rounded to two digits after the point.}\\
\end{tabular}
\end{center}
\label{tab:example1}
\end{table}

The results are summarized in \tabcite{tab:simulations1} and \tabcite{tab:simulations2}.
In all investigated scenarios 
model selection by CAR
scores  is competitive with elastic net regression, and typically
 outperforms the lasso and OLS with no variable selection and OLS
with partial correlation. It is also in most cases
distinctively better than boosting.  Genizi's variable selection criterion also performs
very well, with a similar performance to CAR scores in many cases, except for
Example 2.   \tabcite{tab:simulations1} and \tabcite{tab:simulations2}
also show the true and false positives for each method.
The regression models selected by the CAR score approach often exhibt
the largest number of true positives and the smallest number of false positives,
which explains its effectiveness.

\figcite{fig:m3beta} shows 
the distribution of
the estimated regression coefficients for the investigated methods
over the 200 repetitions for Example 3 with $n=50$ and $\sigma=3$.
This figure demonstrates that  using CAR scores --- unlike lasso, elastic net, and boosting ---
recovers the regression coefficients of variables $X_6$ to $X_{10}$ 
that have negative signs.  Moreover, in this setting the CAR score 
regression coefficients have a much smaller variability than those obtained
using the OLS-Genizi method.

The simulations for Examples 1 and 2 represent cases
where the null variables $X_3$, $X_4$, $X_6$, $X_7$, and
$X_8$ are correlated with the non-null variables $X_1$, $X_2$
and $X_5$.  In such a setting the variable importance $\phi^{\text{CAR}(X_j)}$ assigned
by squared CAR scores to the null-variables is non-zero.
For illustration, we list in \tabcite{tab:example1}
the population quantities for Example 1 with $\sigma=3$.
The squared
multiple correlation coefficients is $\Omega^2 = 0.70$ and the ratio
of signal variance to noise variance equals $\Omega^2/(1-\Omega^2) = 2.36$.
Standardized regression coefficients
$\bb_{\text{std}}$,
as well as partial correlations
$\tilde\bP_{\bX Y}$
are zero whenever the corresponding regression coefficient 
$\bb$ vanishes.
In contrast, 
marginal correlations
$\bP_{\bX Y}$, CAR scores
$\bomega$
and the variable importance
$\phi^{\text{CAR}}(X_j)$
are all non-zero even for $b_j = 0$.
This implies that for large sample size in the setting of Example~1
all variables (but in particular also $X_3$, $X_4$, and $X_6$)
carry information about the response, albeit only weakly and
indirectly for variables with $b_j=0$.

In the literature on variable importance the axiom of
``proper exclusion'' is frequently encountered, i.e.\ it is demanded that
the share of $\Omega^2$ allocated to a variable
$X_j$ with  $b_j=0$ is zero \citep{Groe2007}.
The squared CAR scores violate this principle if null and non-null
variables are correlated.  However, in our view this violation
makes perfect sense, as in this case the null variables are
informative about $Y$ and thus may be useful for prediction.  Moreover, 
 because of the existence of equivalence classes in
graphical models one can construct an alternative regression model 
with the same fit to the data that shows no correlation between null 
and non-null variables but which then necessarily 
includes additional variables.  A related argument against
proper exclusion is found in \citet{Groe2007}.

%\afterpage{\clearpage}

\subsection{Diabetes data}

% definitions for the two data sets
\newcommand\lpsa{{\tt lpsa}}
\newcommand\lcavol{{\tt lcavol}}
\newcommand\lweight{{\tt lweight}}
\newcommand\age{{\tt age}}
\newcommand\lbph{{\tt lbph}}
\newcommand\svi{{\tt svi}}
\newcommand\lcp{{\tt lcp}}
\newcommand\gleason{{\tt gleason}}
\newcommand\pgg{{\tt pgg45}}
\newcommand\sex{{\tt sex}}
\newcommand\bmi{{\tt bmi}}
\newcommand\bp{{\tt bp}}
\newcommand{\ts}[1]{{\tt s#1}}

Next we reanalyzed a low-dimensional benchmark data set on the disease progression 
of diabetes discussed in \citet{EHJT2004}.  
There are $d=10$ covariates, age (\age), sex (\sex), body mass index (\bmi),
blood pressure (\bp) and six blood serum measurements
(\ts{1}, \ts{1}, \ts{2} \ts{3} , \ts{4}, \ts{5}, \ts{6}),
on which data were collected from   $n=442$ patients.
As $d < n$ we used empirical estimates of CAR scores and ordinary least
squares regression
coefficients in our analysis.  The data were centered and standardized
beforehand.

\begin{figure}[p]
\begin{center}
\centerline{\includegraphics[width=1\textwidth]{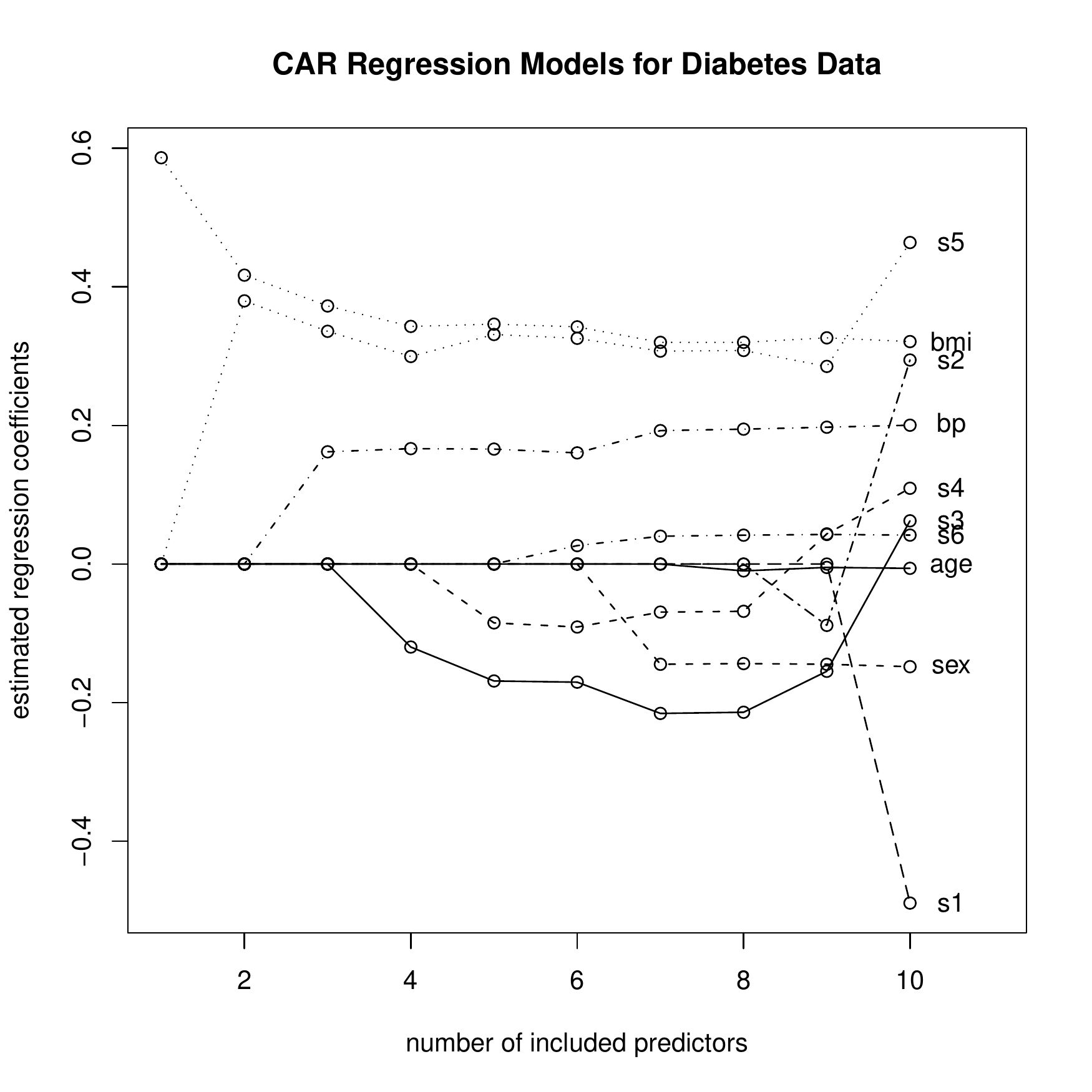}}
\caption{Estimates of regression coefficients for the diabetes study.  Variables
 are included in the order of empirical squared CAR scores, and the
corresponding regression coefficients
are estimated by ordinary least squares.  The antagonistic correlated
variables \ts{1} and \ts{2} are included only in the last two steps.}
\label{fig:diabetes}
\end{center}
\end{figure}

A particular challenge of the diabetes data set is that it contains two variables
(\ts{1} and \ts{2}) that are highly positively correlated but behave in an antagonistic fashion.
Specifically, their regression coefficients have the opposite signs 
so that in prediction the two variables cancel each other out.
\figcite{fig:diabetes} shows all regression models that arise when 
covariates are added to the model in the order of decreasing
variable importance given by $\phi^{\text{CAR}}(X_j)$.  As can be seen from
this plot, the variables $\ts{1}$ and $\ts{2}$ are ranked least important
and included only in the two last steps.

\begin{table}[t]
\caption{Ranking of variables and selected models (in bold type) using various
variable selection approaches on the diabetes data.}
\begin{center}
\begin{tabular}{lrrrrrr}
\toprule
 Rank         & $\tilde \bP_{\bX Y}  $ ${}^*$ &  $\bP_{\bX Y}  $  ${}^*$& CAR ${}^*$  & Elastic Net & Lasso & Boost\\ 
\midrule
\midrule
\age    & 10      & {\bf 8} & 8       & {\bf 10} & ---     & ---    \\
\sex    & {\bf 4} & 10      & 7       & {\bf 4}  & {\bf 5} & {\bf 5}\\
\bmi    & {\bf 1} & {\bf 1} & {\bf 1} & {\bf 1}  & {\bf 1} & {\bf 1}\\
\bp     & {\bf 2} & {\bf 3} & {\bf 3} & {\bf 3}  & {\bf 3} & {\bf 3}\\
\ts{1}  & 5       & {\bf 7} & 9       & {\bf 9}  & {\bf 6} & {\bf 6}\\
\ts{2}  & 6       & {\bf 9} & 10      & {\bf 7}  & ---     & ---    \\
\ts{3}  & 9       & {\bf 5} & {\bf 4} & {\bf 5}  & {\bf 4} & {\bf 4}\\
\ts{4}  & 7       & {\bf 4} & {\bf 5} & {\bf 6}  & ---     & ---    \\
\ts{5}  & {\bf 3} & {\bf 2} & {\bf 2} & {\bf 2}  & {\bf 2} & {\bf 2}\\
\ts{6}  & 8       & {\bf 6} & {\bf 6} & {\bf 8}  & {\bf 7} & {\bf 7}\\
\midrule
Model size  & 4   & 9       &   6     &  10      & 7 & 7 \\
\bottomrule 
\multicolumn{6}{c}{${}^*$ empirical estimates.}\\
\end{tabular}
\end{center}
\label{tab:diabetes}
\end{table}

For the empirical estimates the
exact null distributions are available, therefore we also computed
$p$-values for the estimated CAR scores, 
 marginal correlations $\bP_{\bX Y}$ and partial correlations $\tilde \bP_{\bX Y}$,
and selected those variables for inclusion with a $p$-value
smaller than 0.05.  In addition, we computed lasso, elastic
net and boosting regression models.  

The results are summarized in \tabcite{tab:diabetes}.
All models include \bmi, \bp{}  and \ts{5} and thus agree that those three
explanatory variables are most important for prediction of diabetes progression.
Using marginal correlations and the elastic net both lead to large models
of size 9 and 10, respectively, whereas the CAR feature selection in accordance
with the simulation study results in a smaller model.  The CAR model
and the model determined by partial correlations are the only ones not including
either of the variables $\ts{1}$ or $\ts{2}$.

In addition, we also compared CAR models selected by the various
penalized RSS approaches. Using the $C_p$ / AIC rule on the empirical CAR scores
results in 8 included variables, RIC leads to 7 variables, and BIC
to the same 6 variables as in~\tabcite{tab:diabetes}.

\subsection{Gene expression data}

\begin{table}[!t]
\caption{Cross-validation prediction errors resulting from regression models for
the gene expression data.}
\begin{center}
\begin{tabular}{lr}
\toprule
Model (Size)    & Prediction error\\ 
\midrule
\midrule
Lasso (36)        & 0.4006 (0.0011) \\
Elastic Net (85)  & 0.3417 (0.0068) \\
CAR (36) ${}^*$   & 0.3357 (0.0070) \\
CAR (60) ${}^*$   & 0.3049 (0.0064) \\
CAR (85) ${}^*$   & 0.2960 (0.0059) \\
\bottomrule 
\multicolumn{2}{c}{${}^*$ shrinkage estimates.}\\
\end{tabular}
\end{center}
\label{tab:aging}
\end{table}

Subsequently,
we analyzed data from a gene-expression study  investigating the 
relation of aging and gene-expression in the human frontal cortex
\citep{LP+2004}.  Specifically, the age $n=30$ patients
was recorded, ranging from 26 to 106 years, and the expression
of $d=12\,625$ genes was measured by microarray technology.
In our analysis we used the age as metric response  $Y$ and
the genes as explanatory variables $\bX$.  Thus, our aim 
was to find genes that help to predict the age of the patient.

In preprocessing
we removed genes
with negative values and log-transformed the expression values
of the remaining $d=11\,940$ genes.  We centered and standardized
the data and computed empirical marginal correlations.  Subsequently, 
based on marginal correlations we filtered
out all genes with local false non-discovery rates (FNDR) smaller than 0.2,
following  \citet{AS2010}.  Thus, in this prescreening step 
we retained the $d=403$ variables with  local false-discovery rates smaller than 0.8.

\begin{figure}[ht]
\begin{center}
\centerline{\includegraphics[width=1\textwidth]{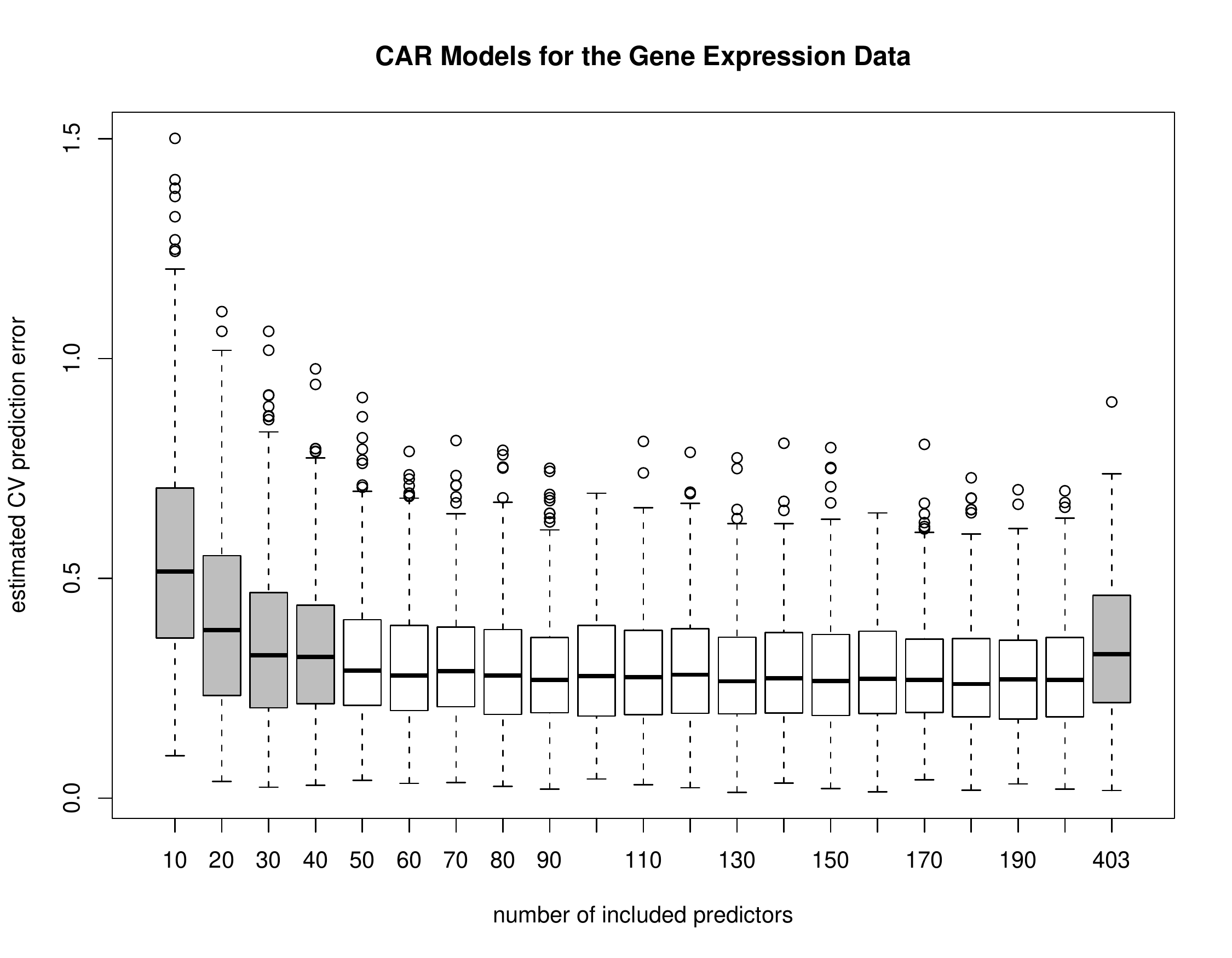}}
\caption{Comparison of CV prediction errors of CAR regression models
of various sizes for the gene expression data.}
\label{fig:carmodels}
\end{center}
\end{figure}

On this $30 \times 403$ data matrix we fitted regression models using shrinkage CAR, lasso, and elastic
net. The optimal tuning parameters
were selected by minimizing prediction error estimated by 5-fold cross-validation
with 100 repeats. Cross-validation  included model selection as integrative
step, e.g., CAR scores were recomputed in each repetition in order
to avoid downward bias. A summary of the results is found in \tabcite{tab:aging}.
The prediction error of the elastic net regression model is substantially
smaller than that
of the lasso model, at the cost of 49 additionally included covariates.  The regression
model suggested by the CAR approach for the same model sizes improves over
both models.  
As can be seen from \figcite{fig:carmodels}
the optimal CAR regression model has a size of about 60 predictors.
The inclusion of additional explanatory variables does not substantially improve
prediction accuracy.

%\clearpage

%\newpage
\section{Conclusion}

We have proposed correlation-adjusted marginal 
correlations $\bomega$, or CAR scores, as a means 
of assigning variable importance to individual predictors
and to perform variable selection.
This approach is based on simultaneous orthogonalization of 
the covariables by Mahalanobis-decorrelation
and subsequently estimating the remaining correlation between the
response and the sphered predictors. 

We have shown that CAR scores not only simplify the regression equations but
more importantly result in a canonical ordering of variables that provides
the basis for a simple yet highly effective  procedure for variable selection.
Because of the orthogonal compatibility of
squared CAR scores they can also be used to assign variable importance 
to groups of predictors.
In simulations and by analyzing experimental data we have shown that
CAR score regression is competitive in terms of prediction and model error
 with regression approaches such as elastic net, lasso or boosting.

Since writing of this paper in 2010 we have now also become aware of
the ``tilted correlation'' approach to variable selection \citep{CF2011}.
The tilted correlation  --- though not identical to the CAR score --- 
 has the same objective, 
namely to provide a measure of the contribution of each covariable in 
predicting the response while taking account of the correlation among explanatory variables.

In summary, as exemplified in our analysis we  suggest the 
following strategy for analyzing high-dimensional data,
using CAR scores for continuous and CAT scores for categorical response:
\begin{enumerate}
\item Prescreen predictor variables using marginal correlations (or $t$-scores)
      with an adaptive threshold determined, e.g., by controlling FNDR \citep{AS2010}. 
\item Rank the remaining variables by their squared CAR (or CAT) scores.
\item If desired, group variables and compute grouped CAR (or CAT) scores.
\end{enumerate}
Currently, we are studying algorithmic improvements  to enable
shrinkage estimation 
of CAT and CAR scores even for very large numbers of predictors and 
correlation matrices, which may render unnecessary in many cases the prescreening step above.

\section*{Acknowledgments}
We thank Bernd Klaus and Carsten Wiuf
for critical comments and helpful discussion.
Carsten Wiuf also pointed out special
properties of the Mahalanobis transform.
Part of this work was supported by BMBF grant  no. 0315452A (HaematoSys project).

%\newpage

\bibliographystyle{apalike}
\bibliography{preamble,econ,genome,stats,array,sysbio,misc,molevol,med,entropy}

\end{document}